\documentclass[aps,pra,twocolumn,groupedaddress,showpacs]{revtex4-1}
\usepackage[T1]{fontenc} %Allows for special characters in the text: ö,é,à etc.
\usepackage{lmodern}
\usepackage{amsmath}
\usepackage[pdftex]{graphicx}
\usepackage{color}
\begin{document}

% Use the \preprint command to place your local institutional report
% number in the upper righthand corner of the title page in preprint mode.
% Multiple \preprint commands are allowed.
% Use the 'preprintnumbers' class option to override journal defaults
% to display numbers if necessary
%\preprint{}

%Title of paper
\title{Optical injection in semiconductor ring lasers}

% repeat the \author .. \affiliation  etc. as needed
% \email, \thanks, \homepage, \altaffiliation all apply to the current
% author. Explanatory text should go in the []'s, actual e-mail
% address or url should go in the {}'s for \email and \homepage.
% Please use the appropriate macro foreach each type of information

% \affiliation command applies to all authors since the last
% \affiliation command. The \affiliation command should follow the
% other information
% \affiliation can be followed by \email, \homepage, \thanks as well.
\author{W. Coomans}
\email[Electronic address: ]{wcoomans@vub.ac.be}
\author{S. Beri}
\altaffiliation{Also with Department of Physics}
\author{G. Van der Sande}
\author{L. Gelens}
\author{J. Danckaert}
\altaffiliation{Also with Department of Physics}
%\homepage[]{Your web page}
%\thanks{}
%\altaffiliation{}
\affiliation{Department of Applied Physics and Photonics, Vrije Universiteit Brussel, Pleinlaan 2, B-1050 Brussel, Belgium}

%Collaboration name if desired (requires use of superscriptaddress
%option in \documentclass). \noaffiliation is required (may also be
%used with the \author command).
%\collaboration can be followed by \email, \homepage, \thanks as well.
%\collaboration{}
%\noaffiliation

\date{\today}

\begin{abstract}
We theoretically investigate optical injection in semiconductor ring lasers and disclose several dynamical regimes. Through numerical simulations and bifurcation continuation, two separate parameter regions in which two different injection-locked solutions coexist are revealed, in addition to a region in which a frequency-locked limit cycle coexists with an injection-locked solution. Finally, an anti-phase chaotic regime without the involvement of any carrier dynamics is revealed. Parallels are drawn with the onset of chaos in the periodically forced Duffing oscillator.
\end{abstract}

% insert suggested PACS numbers in braces on next line
\pacs{42.55.Px, 42.65.Sf, 42.60.Mi}
% insert suggested keywords - APS authors don't need to do this
%\keywords{}

%\maketitle must follow title, authors, abstract, \pacs, and \keywords
\maketitle

% body of paper here - Use proper section commands
% References should be done using the \cite, \ref, and \label commands
% Put \label in argument of \section for cross-referencing
%\section{\label{}}
\section{Introduction}
Optically injected laser systems generally consist of two laser sources, a "master" laser whose output light is coupled into the cavity of a second "slave" laser. A simple model for this type of system is a nonlinear oscillator (slave) which is periodically driven (master). Although these systems are relatively simple, they exhibit a wealth of dynamical behavior which has been widely studied for different types of lasers \cite{Wieczorek_PhysRep_2005,Wieczorek_OptComm_1999,Tartwijk95,Kovanis_OptComm_1999,Pan_APL_1993,Simpson_QSO_1997,Yeung_PRE_1998,ALodiChaos,Gatare_PRE_2009,Osborne_PRA_2009}.

A class of semiconductor lasers for which the nonlinear dynamics induced by optical injection have not yet been investigated in depth are the semiconductor ring lasers (SRLs). A SRL is a semiconductor laser in which the light is confined in a circular waveguide structure. As a result, SRLs generate light in two opposite directions referred to as the clockwise (CW) and the counterclockwise (CCW) mode (see Figure \ref{fig:setup}). SRLs have received increasing attention in recent years \cite{KraussSRL}, because they are suitable candidates as key components in photonic integrated circuits \cite{HillNature}. The bistable character of their directional mode operation allows them to be used in systems for all-optical switching and as all-optical memories \cite{AlmeidaOptSw,HillNature}.
Optical injection can be particularly important in SRLs as a control mechanism for the dynamics in optical switching applications \cite{Perez_OptExpr_2007,Lendert08}, or when a holding beam is used to enforce unidirectional operation in the injected direction \cite{Yuan_Wang_IEEEJQE_14_2008}. However, optical injection can also give rise to very intricate dynamics, which may obstruct the desired dynamical behavior. 
The particular Z$_2$-symmetry of the SRL and its resulting phase space structure has already led to results such as alternative switching mechanisms \cite{Lendert08,LendertPhSpSw}, multistable regimes \cite{MultiStab} and non-Arrhenius mode-hopping \cite{Beri_PRL_2008}. The different dynamical regimes resulting from forcing the SRL through optical injection will be disclosed in this paper.

Given the interest in SRLs, we will investigate the behavior of this two-mode device when subjected to unidirectional optical injection. Optical injection in other two-mode semiconductor lasers such as VCSELs \cite{Gatare_PRE_2009} and two-color lasers \cite{Osborne_PRA_2009} has also recently been investigated. The bimodal character of these devices gives rise to more intricate dynamical behavior, such as bistable operating regimes governed by optical injection. 

In this paper we will perform an extensive bifurcation analysis using the software continuation package AUTO \cite{auto}, complemented by numerically solving the rate equation model. This approach allows us to compute the bifurcation diagrams for optically injected SRLs and to reveal the stability of the invariant structures in the phase space of the mathematical model which will be introduced in the next section.
Keeping in mind the all-optical memory and switching applications, we focus on unidirectional optical injection. This means we optically inject only one of the two counterpropagating modes, chosen to be the CW mode. This will also facilitate comparison of our results with future experiments. We will further assume that the SRL is biased in the bistable unidirectional regime.

This paper is organized as follows. The rate equation model of the optically injected SRL is described in Section \ref{sec:Model}, where we also point out the differences between the SRL model and that of other two-mode lasers.
The analysis starts in Section \ref{sec:Numerics}, in which we reveal some of the characteristic behavior of the optically injected SRL obtained by numerically solving the rate equation model.
In Section \ref{sec:BifAnalysis}, we complement this analysis by presenting the bifurcation curves of stationary points, and point out the differences with other optically injected semiconductor lasers. This allows us to give a bird's eye view of the dynamics exhibited by the optically injected SRL for different values of the detuning and injection powers.
For a certain parameter range, the optically injected SRL exhibits a novel anti-phase chaotic regime, described in Section \ref{sec:Chaos}.
We finally draw conclusions of our analysis and point to future work in Section \ref{sec:Conclusion}.

\section{Formulation of the model\label{sec:Model}}
\begin{figure}
\includegraphics[width=0.8\columnwidth]{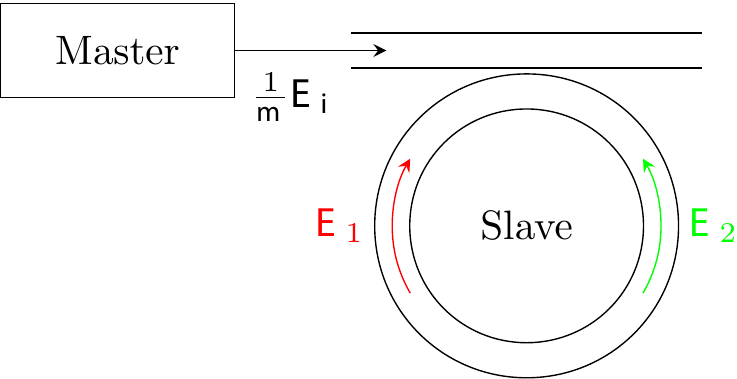}
\caption{(Color online) Schematic setup of the optically injected SRL ($m$ is assumed to be the amplitude coupling coefficient of the straight output waveguide into the ring cavity).\label{fig:setup}}
\end{figure}

We consider a typical master-slave setup in which we neglect coupling from the slave to the master laser (see Figure \ref{fig:setup}). In this setup, the SRL is assumed to operate in a single transverse and single longitudinal mode and can sustain two counterpropagating directional modes.
Following Refs. \cite{SorelAO,Sorel2003} with a straightforward modification to account for the optical injection as in \cite{Lendert08}, we can write the following rate equations for an optically injected SRL; neglecting spatial variations within the laser and adiabatically eliminating the medium's polarization dynamics:
\begin{subequations}
\label{eq:RateEq}
\begin{align}
\frac{\mathrm{d}E_1}{\mathrm{d}t}  &= \kappa \left( {1 + i\alpha } \right)\left[ N\left(1 - s\left|E_1 \right|^2  - c\left| E_2 \right|^2 \right) - 1\right]E_1 \notag \\ 
& \qquad - ke^{i\phi_k}E_2 -i\Delta E_1+ \frac{1}{{\tau _{in}}}E_i, \\
\frac{\mathrm{d}E_2}{\mathrm{d}t}  &= \kappa \left( {1 + i\alpha } \right)\left[ N\left(1 - s\left| 	E_2 \right|^2 - c\left| E_1 \right|^2 \right) - 1\right]E_2 \notag \\
& \qquad - ke^{i\phi_k}E_1 -i\Delta E_2,
\\
	\frac{\mathrm{d}N}{\mathrm{d}t} &= \gamma \left[ \mu - N - N(1 - s\left|E_1 \right|^2 - c\left|E_2 \right|^2 )\left|E_1 \right|^2 \right. \notag \\ 
	& \qquad \left. - N(1 - s\left|E_2  \right|^2 - c\left|E_1 \right|^2 )\left|E_2 \right|^2\right].
\end{align}
\end{subequations}
Here $t$ is time, $E_1$ and $E_2$ are the slowly varying complex envelopes of the counterpropagating waves, $N$ is the carrier population inversion, $\mu$ is the renormalized injection current ($\mu=0$ at transparency and $\mu=1$ at lasing threshold), $\kappa$ is the field decay rate, $\gamma$ is the carrier decay rate, $\alpha$ is the linewidth enhancement factor and $\tau_{in}$ is the cavity round-trip time. 
The two control parameters are the injected field amplitude $E_i>0$ and its detuning $\Delta$ from the longitudinal mode frequency of the SRL. A detuning $\Delta>0$ corresponds to a higher master than slave frequency.
The two counterpropagating waves saturate both their own and each others gain through spectral hole burning and carrier heating effects. These self- and cross-saturation phenomena occur on faster time scales than the photon lifetime of the SRL \cite{SatTimescale}, allowing them to be added phenomenologically, modeled by $s$ and $c$. Note that the cross-saturation is stronger than the self-saturation ($c\approx 2s$) \cite{Javaloyes_JQE_2009}. In addition to this nonlinear coupling there also exists a linear coupling between the counterpropagating waves, referred to as backscattering. It is caused by reflections inside the cavity at the interface with the coupling waveguide and at the cleaved end facets of the output waveguide. They result in a linear coupling between the two fields, modeled by an amplitude $k$ and a phase shift $\phi_k$.
Finally, note that the reference frame of the equations is chosen to corotate with the phase of the master laser so that fixed points of this system correspond to injection-locked states.

In a typical experimental setup, the photon lifetime $\tau_p=\kappa^{-1}$ and the carrier lifetime $\tau_c=\gamma^{-1}$ are respectively of the orders 10 ps and 5 ns, yielding two different time scales in the system. The other parameters are fixed to realistic values $\alpha=3.5$, $s=0.005$, $c=0.01$, $k=0.4412~\mathrm{ns}^{-1}$, $\phi_k=1.4966$ and $\tau_{in}=0.6$ ps \cite{SorelAO}. The value of the bias current $\mu$ is chosen such that the SRL operates in the bistable unidirectional regime, but still relatively close to the alternate oscillation regime \cite{LendertPhSpSw}. The detuning is varied up to 7 ns$^{-1}$ (angular frequency), while the values used for the injection amplitude $E_i$ span several orders of magnitude, ranging from $O(10^{-7})$ up to $O(10^{-2})$.

For future reference, we will briefly highlight the characteristic timescales of the solitary SRL. The principal time scale in any semiconductor laser is the relaxation oscillation timescale. For our set of equations (\ref{eq:RateEq}), the relaxation oscillation angular frequency can be approximated by
\begin{equation}
 \omega_{\mathrm{R}}\approx \sqrt{2(\mu-1)\gamma\kappa}\approx5.307\mathrm{~ns^{-1}}.
\end{equation}
In the case of SRLs there is also a second important time scale regarding intensity oscillations, the alternate oscillation frequency. It characterizes a particular operating regime of the solitary SRL in which two coupling mechanisms---the cross-gain saturation and the backscattering---compete with each other, inducing intensity oscillations at an angular frequency given by \cite{SorelAO}
\begin{equation}
\omega_{\mathrm{AO}} = 2k\sqrt{-\cos(2\phi_k)-\alpha\sin(2\phi_k)} \approx 0.606 \mathrm{~ns}^{-1}.
\label{eq:FreqAO}
\end{equation}
In the bistable unidirectional regime, the SRL has four different steady state solutions in the absence of optical injection. Two of them are stable quasi-unidirectional solutions with the power concentrated in either the CW or the CCW mode. The other two are unstable bidirectional solutions with equal power in both modes, with the counterpropagating fields respectively in-phase (IP) and out-of-phase (OP). They are each characterized by a particular optical frequency $\omega_\mathrm{X}$, with X=\{CW,CCW,IP,OP\}. This is due to the different carrier densities associated to each of these solutions. The carrier densities are influenced by the optical intensity, in which the backscattering also plays a role by altering the effective gain of both modes. This frequency corresponds to a certain detuning with respect to the cavity resonance frequency $\omega_\mathrm{0}$ which can be calculated numerically and is given by
\begin{align}
\Delta_{\mathrm{CW}}=\Delta_{\mathrm{CCW}}\approx0.225\mathrm{~ns^{-1}}\label{eq:FreqUni}\\
\Delta_{\mathrm{OP}}=-\Delta_{\mathrm{IP}}\approx0.326\mathrm{~ns^{-1}}\label{eq:FreqBi}.
\end{align}
with $\Delta_\mathrm{X}=\omega_\mathrm{X}-\omega_\mathrm{0}$. Note that $\Delta_{\mathrm{CW}}$ must be equal to $\Delta_{\mathrm{CCW}}$ due to the symmetry properties of our system ($E_1$ and $E_2$ may be exchanged). The fact that $\Delta_{\mathrm{OP}}=-\Delta_{\mathrm{IP}}$ follows immediately from the equations (\ref{eq:RateEq}).

Contrary to two-color lasers where the mode spacing is highly nondegenerate  \cite{Osborne_PRA_2009}, the mode spacing in SRLs is degenerate since the counterpropagating modes have identical frequencies. For this reason, the counterpropagating modes in SRLs have a significant phase coupling.  On the other hand, the phases of the two modes in two-color lasers are decoupled and only the intensity of the uninjected mode influences the dynamics \cite{Osborne_PRA_2009}.
Moreover, contrary to SRLs, saturation effects in two-color lasers are such that self-saturation is stronger than cross-saturation \cite{Osborne_PRA_2009}, affecting the relative stability of the modes.
Both the SRL model and the spin-flip model for VCSELs \cite{SanMiguel_PRA_1995,MartinRegalado_JQE_1997} account for a phase coupling between the modes. However, in the spin-flip model, the optical injection studied in Ref. \cite{Gatare_PRE_2009} corresponds to injection in both modes when the VCSEL is lasing in both circular modes simultaneously. Finally, the SRL model does not need an extra dimension for a second carrier population as the VCSEL model does.

\begin{figure}[t!]
	\centering
		\includegraphics[width=\columnwidth]{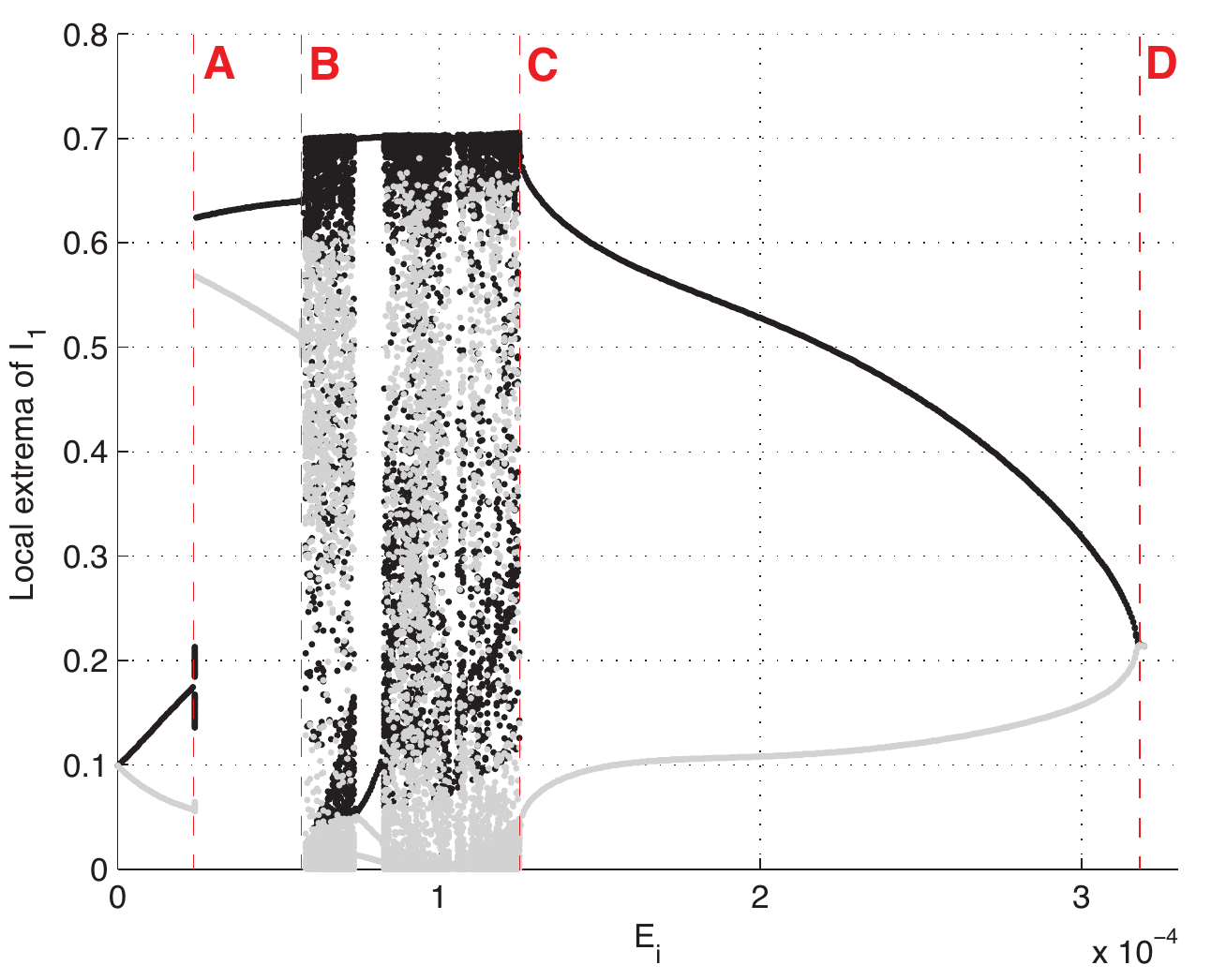}
	\caption{(Color online) Orbit diagram depicting the local extrema of the intensity $|E_{1}|^{2}$ of the CW mode versus the injection amplitude $E_i$. Black (gray) dots indicate local maxima (minima).
$\mu$ = 1.704, $\phi_k$ = 1.4966, $k$ = 0.4412 ns$^{-1}$, $\alpha$ = 3.5, $c$ = 0.01,
$s$ = 0.005, $\kappa$ = 100 ns$^{-1}$, 
$\gamma$ = 0.2 ns$^{-1}$, $\Delta$ = -1 ns$^{-1}$.\label{fig:OrbitDiagram1}}
\end{figure}

\section{Numerical simulations\label{sec:Numerics}}
We start our analysis by numerically solving (\ref{eq:RateEq}) for different values of the detuning $\Delta$ and the injected field amplitude $E_i$, using a fourth order Runge-Kutta algorithm with a fixed time step of 1 ps. We have constructed orbit diagrams for different values of the detuning using $E_{i}$ as a parameter.
These orbit diagrams plot the local extrema of the system's attractor as a function of the injection amplitude (or only a part of the attractor if its basin of attraction is not the whole phase space).

The orbit diagram for $\Delta=-1\mathrm{~ns}^{-1}$, only depicting the local extrema of the intensity of the CW mode, is shown in Figure \ref{fig:OrbitDiagram1}. For low injection amplitudes the output power starts to oscillate around the former steady state, due to the beating between the injected signal and the optical fields inside the SRL. When increasing $E_i$, the amplitude of the oscillation grows and eventually the SRL switches mode (at point A) due to the optical injection in the CW mode (the solitary SRL was assumed to reside in the CCW mode prior to the optical injection).

\begin{figure}
	\centering
		\includegraphics[width=\columnwidth]{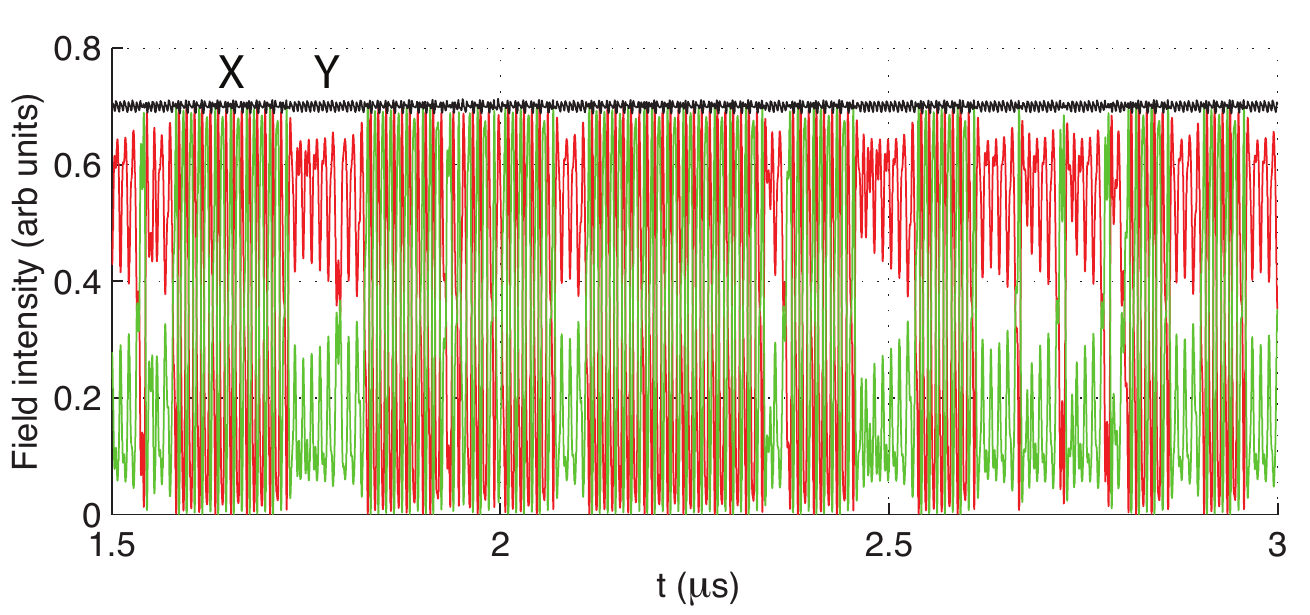}
	\caption{(Color online) Representative time trace of the respective intensities of the CW (dark gray, red online) and the CCW mode (light gray, green online), and the total intensity (black) inside the cavity of the SRL when operating in the chaotic regime. Transient behavior has been eliminated. $\mu$ = 1.704, $\phi_k$ = 1.4966, $k$ = 0.4412 ns$^{-1}$, $\alpha$ = 3.5, $c$ = 0.01,
$s$ = 0.005, $\kappa$ = 100 ns$^{-1}$, 
$\gamma$ = 0.2 ns$^{-1}$, $E_i$ = 6 10$^{-5}$, $\Delta$ = -1 ns$^{-1}$.\label{fig:ChaosTT}}
\end{figure}

At point B the attractor changes from a limit cycle to a more complicated structure, yielding a sudden burst of local extrema. The time trace corresponding to this particular region is shown in Figure \ref{fig:ChaosTT}.
In Section \ref{sec:Chaos}, we will later show that it corresponds to a chaotic regime. However, this regime has a different origin than those observed in other optically injected laser systems. From point B up to point C, the orbit diagram reveals chaos interspersed with periodic windows.

For higher $E_i$ (C$\rightarrow$D), the system relaxes to a stable limit cycle which eventually dies out in a Hopf bifurcation at point D. At that point, the SRL locks to the injected signal. 
In the next section, we will reveal that this particular injection-locked solution is only one out of three different injection-locked solutions (the bidirectional one). 
The other two (unidirectional) injection-locked solutions have the same route to locking as the regular injection problem, which happens through a saddle-node bifurcation for low injection powers \cite{Wieczorek_OptComm_1999,Kovanis_OptComm_1999,Yeung_PRE_1998}.

Figure \ref{fig:OrbitDiagram1} shows a Hopf route to locking of the bidirectional injection-locked solution. There is also a Hopf route to locking in other optically injected semiconductor lasers  \cite{Kovanis_OptComm_1999}, but it only occurs for much higher injection powers. In that case, considering that for a fixed value of the detuning one would continuously raise the injection power from zero, the slave laser would first injection-lock through a saddle-node bifurcation, after which it unlocks because of the undamping of relaxation oscillations and finally locks again through a Hopf bifurcation. In our scenario, the very first locking event happens through a Hopf bifurcation.

\section{Bifurcation analysis\label{sec:BifAnalysis}}
\begin{figure}
\includegraphics[width=\columnwidth]{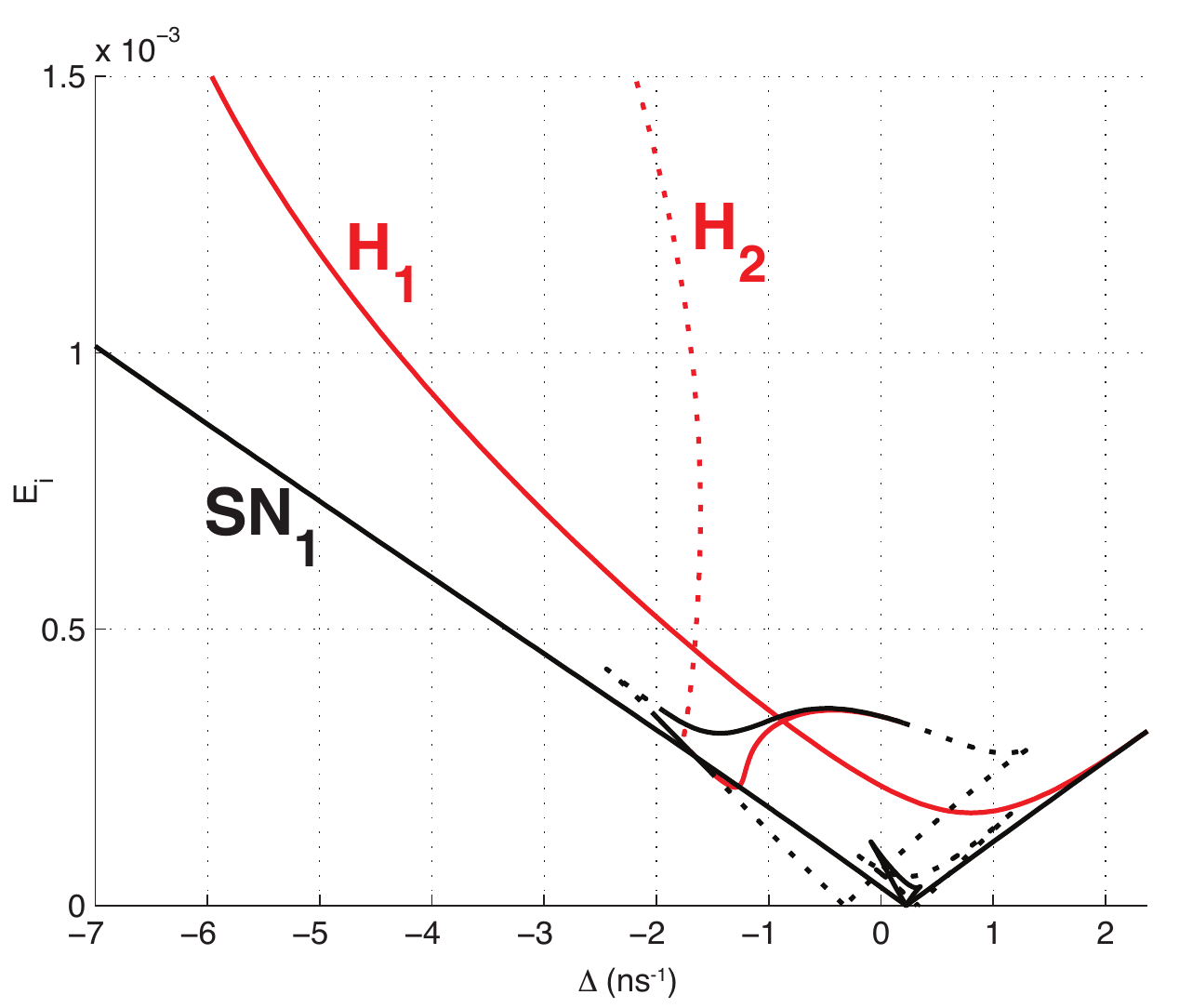}
\caption{(Color online) Bifurcation diagram in the $(\Delta,E_i)$-plane, generated using the rate equations (\ref{eq:RateEq}). Saddle-node (Hopf) bifurcations are depicted in black (light gray, red online). Supercritical (subcritical) bifurcations are depicted in full (dashed) lines.
\label{fig:InjLockingDiagram}}
\end{figure}

The dynamical behavior of the solutions of (\ref{eq:RateEq}) will generally vary for different values of the injected field amplitude $E_i$ and the detuning $\Delta$. Qualitative changes in the dynamical behavior of the system, so-called bifurcations, can be numerically detected and continued in this two-dimensional parameter space using for example the bifurcation continuation package AUTO \cite{auto}. Figure \ref{fig:InjLockingDiagram} shows different regions in the ($\Delta$,$E_i$)-plane bounded by bifurcation lines, with each region corresponding to different dynamical behavior.

Two of these bifurcation lines, SN$_1$ and H$_1$, are familiar. They arise in many other optically injected lasers for small injection amplitude and detuning \cite{Annovazzilodi,Wieczorek_OptComm_1999,Wieczorek2002,Yeung_PRE_1998,Tartwijk95}. The SN$_1$-line represents a saddle-node bifurcation on a cycle (infinite-period bifurcation), while the H$_1$-line represents a Hopf bifurcation. The region confined between these two lines is the stable locking region, where the SRL is phase-locked to the injected signal, yielding an injection-locked solution $s_{cw}$. The transformation of this solution near the boundaries of the stable locking region is identical to other optically injected laser systems \cite{Wieczorek2002,Wieczorek_OptComm_1999}. For injection powers higher than H$_1$, the SRL exhibits intensity oscillations at approximately the relaxation oscillation frequency $\omega_R$. 
For injection powers just below SN$_1$, the SRL also exhibits a periodic solution which lengthens its oscillation period when raising the injection power. This period becomes infinite when crossing SN$_1$, where it locks to the injected signal.

\begin{figure}
\includegraphics[width=\columnwidth]{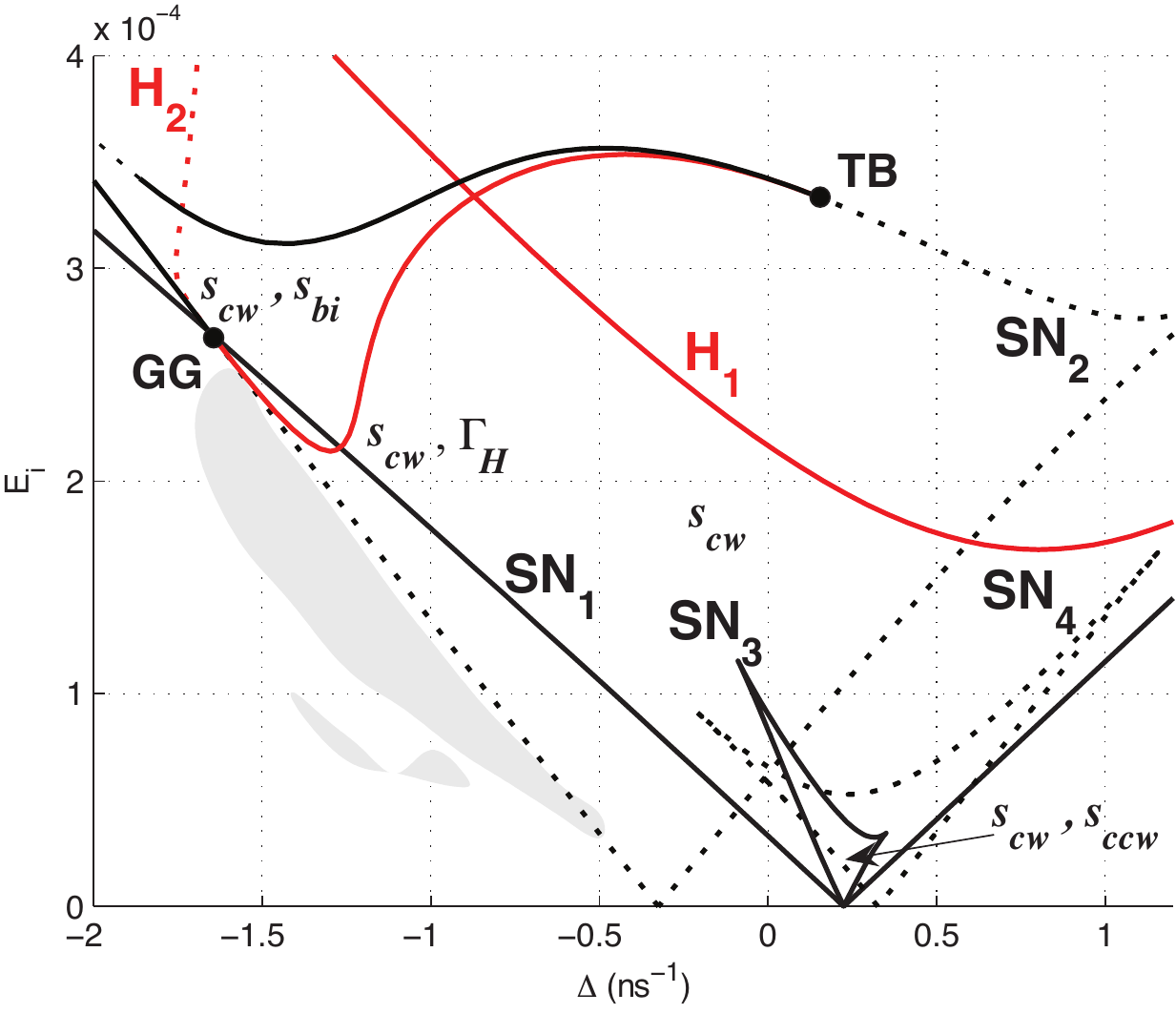}
\caption{(Color online) Bifurcation diagram in the $(\Delta,E_i)$ plane. This is a blow-up of Figure \ref{fig:InjLockingDiagram} in order to show the three new saddle-node bifurcations (SN$_2$, SN$_3$ and SN$_4$), which correspond to three additional resonance tongues.
The curves SN$_2$ and SN$_4$ have their origin symmetrically around $\Delta=0$ (originating from the bidirectional solutions), while the SN$_3$ curve has the same origin as the SN$_1$ curve of Figure \ref{fig:InjLockingDiagram} (originating from the unidirectional solutions). The notations $s_{cw}$, $s_{ccw}$, $s_{bi}$ and $\Gamma_H$ indicate where these respective solutions (co)exist.
The shaded area indicates regions of chaotic behavior. Conventions as in Figure \ref{fig:InjLockingDiagram}.
\label{fig:DiagramSNZoom}}
\end{figure}

So far standard optical injection behavior has been described. However, the bifurcation diagram in Figure \ref{fig:InjLockingDiagram} reveals the presence of bifurcation lines which are not present for a single-mode semiconductor laser \cite{Wieczorek_OptComm_1999}. The H$_2$ line is a Hopf bifurcation which has a different origin than the one encountered in VCSELs \cite{Gatare_PRE_2009} and two-color lasers \cite{Osborne_PRA_2009}; while the SN$_2$, SN$_3$ and SN$_4$ lines in Figure \ref{fig:DiagramSNZoom} are saddle-node bifurcations, which to the best of our knowledge have not been encountered in other optically injected laser systems. The H$_2$ Hopf bifurcation is supercritical for values of the detuning close to zero and subcritical for more negative values of the detuning (see Figure \ref{fig:DiagramSNZoom}). The stable periodic solution $\Gamma_H$ associated to the supercritical part of H$_2$ disappears when crossing the H$_2$-line upwards, where it turns into an injection-locked steady state $s_{bi}$. This new injection-locked solution $s_{bi}$ is different from $s_{cw}$ which is generated in the saddle-node bifurcation. Both solutions are phase-locked to the master laser, but their power distribution amongst the counterpropagating modes differs; $s_{cw}$ has the optical power concentrated in the mode in which we optically inject (CW), while $s_{bi}$ has approximately equal powers in both modes. In Figure \ref{fig:DiagramSNZoom} it can be seen that there is a parameter region in which $s_{cw}$ and $s_{bi}$ coexist.
However, when raising the injection power, $s_{bi}$ disappears at the SN$_2$ line while $s_{cw}$ only disappears at the H$_1$ line. For certain detunings, the H$_2$ line is located at lower injection power than the SN$_1$ line, yielding a slightly earlier injection-locking ($s_{bi}$ appears at lower injection power than $s_{cw}$).

In the limit cycle $\Gamma_H$ the phase of the laser field is bounded. This means that the phase variables $\phi_1$ and $\phi_2$ are trapped inside a $2\pi$-wide interval, never crossing its boundaries (as opposed to an unbounded or running phase solution which freely runs around the phase circle). Because the reference frame of our equations is chosen to corotate with the phase of the master laser this implies that although the intensities oscillate, the emitted optical frequency is centralized around the master laser frequency. The SRL fields are frequency-locked to the master laser in $\Gamma_H$ but not phase-locked. The parameter region in which $\Gamma_H$ exists is partly bounded by H$_2$ where it turns into a phase-locked (injection-locked) solution $s_{bi}$. When moving toward more positive values of the detuning $\Gamma_H$ disappears in a global infinite-period bifurcation (which originates from the TB point, see further on), which is not depicted in Figure \ref{fig:DiagramSNZoom}. When lowering the injection power $\Gamma_H$ disappears in a chaotic attractor, approximately at the SN$_2$ line. This transition can be seen in Figure \ref{fig:OrbitDiagram1}. There is a clear overlap between the region where $\Gamma_H$ exists and the region confined between SN$_1$ and H$_1$ where $s_{cw}$ exists, yielding the coexistence of an injection-locked solution and a frequency-locked limit cycle.

The SN$_2$, SN$_3$ and SN$_4$ lines are three resonance tongues. In the case of an optically injected single mode semiconductor laser we are only faced with a single resonance tongue, corresponding to our SN$_1$ line. The occurrence of four different resonance tongues in the case of a SRL can be explained by the presence of four different steady state solutions for the solitary semiconductor ring laser at the bias current value we have chosen \cite{LendertPhSpSw}. In the unidirectional regime the solitary SRL has two stable unidirectional solutions and two unstable bidirectional solutions. The frequencies of these unperturbed solutions, which we calculated in Section \ref{sec:Model}, can now be matched to each resonance tongue since the origin of a resonance tongue is located at the frequency of the original non-injected solution.

For weak injection, the SN$_1$ and SN$_3$ lines are located at a detuning $\Delta=\Delta_{\mathrm{CW}}=\Delta_{\mathrm{CCW}}$ (this can be verified in Figure \ref{fig:DiagramSNZoom}, see eq. (\ref{eq:FreqUni}) for the numerical value), indicating that these bifurcation curves are related to the two unidirectional modes. The SN$_1$ line corresponds to the CW solution in which we inject and yields the stable locking boundary, while the SN$_3$ line corresponds to the CCW solution. In the same way a detuning $\Delta=\Delta_{\mathrm{IP}}$ and $\Delta=\Delta_{\mathrm{OP}}$ (see eq. (\ref{eq:FreqBi})) indicates that the SN$_2$ and SN$_4$ lines respectively correspond to the in-phase and the out-of-phase bidirectional solution. Note that these bidirectional solutions are both unstable for the solitary SRL, so for low injection powers the SN$_2$ and SN$_4$ lines correspond to bifurcations of unstable structures. Nevertheless part of the SN$_2$ line is a bifurcation of a stable structure, more precisely a bifurcation in which the injection-locked solution $s_{bi}$ disappears, as mentioned before.

The presence of the SN$_3$ line confirms the intuitive reasoning that due to the symmetry of the solitary SRL (stable CW and CCW states) and the phase-coupling between the fields, two separate injection-locked states should originate from the CW and the CCW solution at low injection power. When crossing the SN$_3$ line from below, an injection-locked solution $s_{ccw}$ appears near the original CCW solution through a saddle-node bifurcation. It is the CCW equivalent of the $s_{cw}$ solution associated to the SN$_1$ line, but it has a much smaller basin of attraction. The SN$_3$ line is both steeper and truncated at the top compared to the SN$_1$ line. The increased steepness can be understood from the backscattering phenomenon. The optically injected light that effectively couples into the CCW mode does so through backscattering. Therefore, the amount of optically injected light coupled into the CCW mode is smaller than that coupled into the CW mode, increasing the amount of optically injected light needed to phase-lock the CCW mode. For higher injection powers (higher $E_i$), $s_{ccw}$ becomes unstable because of the increased effective gain of the CW mode when optically injecting it.

Finally, the SN$_4$ line is a bifurcation of the out-of-phase bidirectional solution, which is an inherent unstable structure. Crossing the SN$_4$ line does not lead to a readily observable change in the system dynamics.

In Figure \ref{fig:DiagramSNZoom} we can also see that the Hopf curve H$_2$ and the saddle-node curve SN$_2$ become tangent to each other at two different points, so-called codimension-2 bifurcation points. An analysis of the eigenvalues near these points reveals that the TB point above the stable locking region is a Takens-Bogdanov point, yielding a double-zero bifurcation at that point \cite{Kuznetsov}. It follows from the local theory that there must be a global (homoclinic) bifurcation line originating from the TB point somewhere between H$_2$ and the subcritical part of SN$_2$. This bifurcation line is not visible in Figure \ref{fig:DiagramSNZoom}, but it is responsible for the disappearing of the limit cycle $\Gamma_H$, as mentioned before.

The GG point is a Gavrilov-Guckenheimer point, yielding a fold-Hopf (or zero-pair) bifurcation \cite{Kuznetsov}, only involving the SN$_2$ and the H$_2$ bifurcation lines (the SN$_1$ line in Figure \ref{fig:DiagramSNZoom} only seemingly crosses the GG point).
The nature of the GG point, which is located at $\Delta_{\mathrm{GG}}\approx-1.645\mathrm{~ns^{-1}}$, implies that the H$_2$ Hopf bifurcation changes stability when passing GG while increasing $\Delta$ \cite{Kuznetsov}. In our case this means that for $\Delta<\Delta_{\mathrm{GG}}$, the limit cycle created by (and below) H$_2$ is unstable, while for $\Delta>\Delta_{\mathrm{GG}}$ it is stable. Moreover, the unstable cycle originates from the unstable node generated by SN$_2$ (not from $s_{bi}$, from which the stable limit cycle is generated) \cite{Kuznetsov}. This implies that $s_{bi}$ does not disappear when crossing the H$_2$ line for $\Delta<\Delta_{\mathrm{GG}}$, but only when crossing the SN$_2$ line, which is confirmed by the numerical simulations.

\section{Anti-phase chaos\label{sec:Chaos}}
The occurrence of chaotic regimes in optically injected laser systems and photonic integrated circuits is a well-known phenomenon, and has been found both experimentally and by numerical simulation \cite{Wieczorek_PhysRep_2005,Sacher_PRA_1992,Gatare_PRE_2009,Simpson_PRA_1995,Kovanis_APL_1995,ALodiChaos,Yousefi_PRL_2007}. For the optically injected SRL we want to focus on a particular chaotic regime (see Figure \ref{fig:ChaosTT}).
It is located below the stable locking boundary, and at detunings significantly lower than the relaxation oscillation frequency $\omega_\mathrm{R}$. The parameter region in which this behavior can be found is indicated in Figure \ref{fig:DiagramSNZoom} by the shaded area. This parameter region has been constructed by interpolating points obtained by numerical simulation of equations (\ref{eq:RateEq}).
Both the CW and the CCW mode exhibit purely chaotic behavior, but nevertheless act in anti-phase. 
This is illustrated in Figure \ref{fig:ChaosTT} by the approximately constant value of the total optical power inside the SRL (black line).

We can distinguish two qualitatively different regimes, denoted by X and Y in Figure \ref{fig:ChaosTT}. Regime X is an oscillation of the counterpropagating modes at a fundamental frequency $\omega_{AO}$, making it similar to the alternate oscillation regime observed in solitary SRLs (see eq. (\ref{eq:FreqAO})) \cite{SorelAO}. Regime Y is characterized by the suppression of the non-injected mode (CCW) by the injected mode (CW) while oscillating at the double frequency $2\omega_{AO}$. The time that the SRL resides in either of these regimes also seems to be chaotically distributed.

Unlike other chaotic regimes in optically injected semiconductor lasers, the anti-phase chaotic regime in an optically injected SRL shows no involvement of any carrier dynamics. The variation in carrier inversion is less than 0.4 \%, which is consistent with the constant total power in Figure \ref{fig:ChaosTT}. Hence, the chaos is purely due to the bistable character of the SRL and relaxation oscillations do not play a role in the onset of chaos as it is the case in other optically injected laser systems \cite{ALodiChaos,Gatare_PRE_2009,Yousefi_PRL_2007,Kovanis_APL_1995,Simpson_PRA_1995,Sacher_PRA_1992}.

In order to get more insight in the appearance of the strange attractor responsible for the chaotic behavior, we investigate the orbit in a different phase space. Ref. \cite{Guy08} introduces a reduced two-dimensional model for the solitary SRL which is valid on timescales slower than the relaxation oscillations. It consists of a variable $\theta \in [-\pi/2,\pi/2]$, a measure for the distribution of the optical power among the counterpropagating modes, and a phase variable $\psi \in [0,2\pi]$, the phase difference between the counterpropagating CW and CCW waves. A value of $\theta = -\pi/2$ corresponds to unidirectional CW operation, while a value of $\theta = \pi/2$ corresponds to unidirectional CCW operation. The validity of this two-dimensional phase space has been confirmed by experiments on a solitary SRL \cite{MultiStab,Beri_PRL_2008}.

\begin{figure}
	\centering
		\includegraphics[width=\columnwidth]{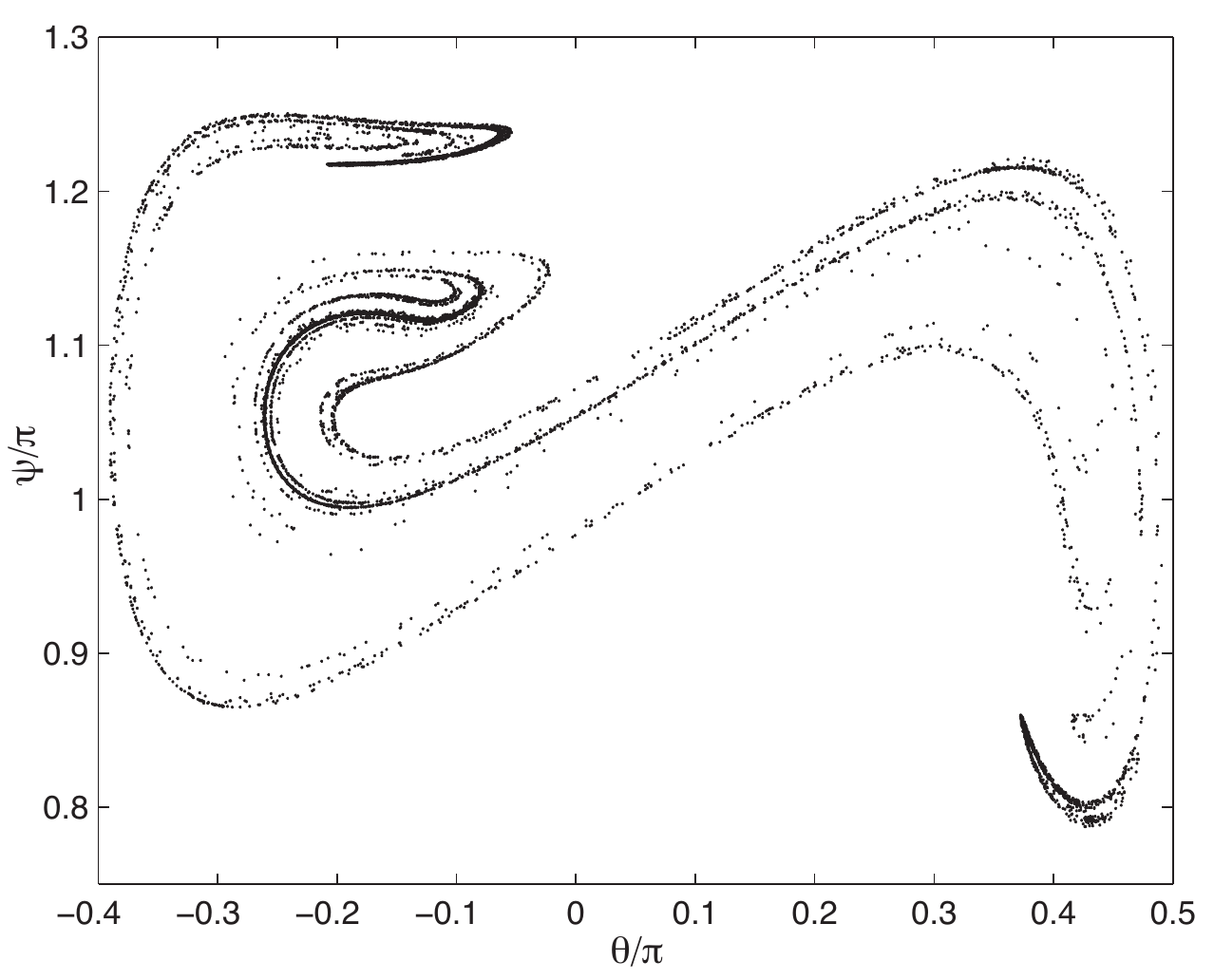}
	\caption{Poincaré section corresponding to an orbit of approximately 100 $\mu$s ($\approx 16000$ points) in the reduced two-dimensional $(\theta,\psi)$ phase space of the solitary SRL, when operating in the regime depicted in Figure \ref{fig:ChaosTT}.\label{fig:PoincareSection}}
\end{figure}

Projecting the five-dimensional chaotic orbit onto this two-dimensional $(\theta,\psi)$ phase plane yields a phase portrait which is hard to interpret. However, when viewed as a Poincaré section, it displays considerable structure. Since the phase $\phi_1$ of the electric field of the CW mode is unbounded in this regime, we construct the Poincaré section by strobing the system whenever $\phi_1$ is a multiple of $2\pi$ (see Figure \ref{fig:PoincareSection}).
The points now fall on a fractal set, which can be interpreted as a cross section of the system's strange attractor in the anti-phase chaotic regime.

Note the topological resemblance of Figure \ref{fig:PoincareSection} to the Poincaré section of the chaotic behavior of a periodically forced Duffing oscillator (see for example Refs. \cite{Strogatz,Guckenheimer}). A Duffing oscillator is a nonlinear oscillator with a cubic stiffness term, which corresponds to weakly damped motion on a double well potential \cite{Guckenheimer}. Hence, its symmetry is very similar to the optically injected SRL. The periodic forcing is a sinusoidal term in the Duffing equation which makes the system nonautonomous, corresponding to the optical injection in the SRL. Furthermore, the double well potential (with two symmetric stable states) has its counterpart in the two stable unidirectional solutions of the solitary SRL. Incited by this resemblance in symmetry properties and strange attractor topology, we conjecture that the way the optically injected SRL evolves to the anti-phase chaos is topologically identical to the way the periodically forced Duffing oscillator evolves to chaos.
The fact that we observe a period doubling of the periodic solution just before the onset of chaos seems to endorse this hypothesis \cite{Guckenheimer}.

Since the periodically forced Duffing oscillator is an extensively studied archetypical dynamical system, we expect that the analogy with the optically injected SRL will allow to interpret and predict dynamical regimes of operation which are a consequence of the device symmetry.

\section{Discussion and concluding remarks\label{sec:Conclusion}}
In this paper, we have theoretically investigated optical injection in SRLs. Starting from a single-longitudinal mode rate equation model for SRLs, we have used numerical simulations and a bifurcation analysis to reveal all the relevant dynamical regimes that unfold for different parameter values. We have focused on optical injection where power is injected in only one of the two counterpropagating modes, as this is the case in multiple experimental setups. The breaking of the intrinsic Z$_2$-symmetry of the SRL due to this optical injection leads to remarkable differences in its dynamical behavior compared to other optically injected laser systems.

Our bifurcation analysis in Section \ref{sec:BifAnalysis} showed that the behavior of one injection-locked solution is similar to the injection-locking of other laser systems. It exists in the stable locking region which is bounded by a saddle-node (for low injected power) and a Hopf bifurcation line (for high injected power). This injection-locked solution has the modal power concentrated in the CW mode in which we inject. Hence, it has a high modal power suppression ratio.
However, the intrinsic bistability of the SRL leads to three separate parameter regions in which this injection-locked state respectively coexists with a bidirectional injection-locked state (which has a lower suppression ratio), a CCW injection-locked state (in which the power is concentrated in the non-injected CCW mode) and a frequency-locked limit cycle.
This frequency-locked limit cycle bifurcates into the bidirectional injection-locked solution through a Hopf bifurcation when raising the injected power.

Our numerical analysis revealed a novel anti-phase chaotic regime at a value of the detuning which is much lower than the relaxation oscillation frequency. It is different from chaotic regimes observed in other optically injected semiconductor lasers because it does not involve any carrier dynamics and the total power emitted by the SRL remains constant. Furthermore, parallels to the onset of chaos in the periodically forced Duffing oscillator will be the subject of future research.

\begin{acknowledgments}
This work has been partially funded by the European Union under project IST-2005-34743 (IOLOS) and by the Research Foundation-Flanders (FWO). This work was supported by the Belgian Science Policy Office under grant IAP-VI10. G.V. and S.B. are Postdoctoral Fellows and L.G. and W.C. are PhD Fellows of the FWO. Furthermore, the authors thank Prof. T. Erneux (Université Libre de Bruxelles) and Dr. I.~A. Khovanov (University of Warwick) for stimulating discussions. The authors also wish to thank the anonymous reviewer for his or her constructive comments on the manuscript.
\end{acknowledgments}

\end{document}